\documentclass[a4paper,11pt]{article}
\pdfoutput=1 

\usepackage{jcappub} 

\usepackage[T1]{fontenc} 
\usepackage{appendix}
\usepackage{bm}
\usepackage{hyperref}

\def\rmd{\mathrm{d}}

\def\ltsima{$\; \buildrel < \over \sim \;$}
\def\gtsima{$\; \buildrel > \over \sim \;$}
\def\simlt{\lower.5ex\hbox{\ltsima}}
\def\simgt{\lower.5ex\hbox{\gtsima}}

\title{The effect of Limber and flat-sky approximations on galaxy weak
  lensing}

\author[a]{Pablo Lemos,}
\author[a,b]{Anthony Challinor,}
\author[a]{and George Efstathiou.}
\affiliation[a]{Institute of Astronomy and Kavli Institute for
  Cosmology Cambridge,\\Madingley Road, Cambridge CB3 OHA, UK}
\affiliation[b]{DAMTP,\\ Centre for Mathematical Sciences,
Wilberforce Road, Cambridge CB3 0WA, UK}

\emailAdd{pl411@cam.ac.uk}
\emailAdd{a.d.challinor@ast.cam.ac.uk}
\emailAdd{gpe@ast.cam.ac.uk}

\abstract{We review the effect of the commonly-used
  Limber and flat-sky approximations on the calculation of shear power
  spectra and correlation functions for galaxy weak lensing. These approximations are accurate at small scales, but it has
  been claimed recently that their impact on low multipoles could lead
  to an increase in the amplitude of the mass fluctuations inferred
  from surveys such as CFHTLenS, reducing the tension between
  galaxy weak lensing and the amplitude determined by Planck from
  observations of the cosmic microwave background. Here, we
  explore the impact of these approximations on cosmological
  parameters derived from weak lensing surveys, using the CFHTLenS
  data as a test case. We conclude that the use of small-angle
  approximations for cosmological parameter estimation is negligible
  for current data,
  and does not contribute to the tension between current weak lensing
  surveys and Planck.\footnote{As this paper was nearing completion,
    analyses were presented by Kilbinger et al.~\citep{Kilbinger:2017} and in the revised version of Kitching et al.~\citep{limits} with similar conclusions to ours.}}


\begin{document}
\maketitle
\flushbottom

\section{Introduction}
\label{sec:intro}

The amplitude and shape of the mass fluctuation spectrum is of
fundamental importance to cosmology. The mass fluctuation spectrum can
be used to test the physics of the early Universe, its contents, 
and the nature of gravity. In addition, it determines the
timescales and evolutionary paths for the formation of non-linear
objects such as galaxies and galaxy clusters that we see in the
Universe today.

Observations of the cosmic microwave background (CMB)
\citep{Planck:2013, Planck:2015} have led to precision measurements of
the shape of the fluctuation spectrum and via gravitational lensing
of the CMB can constrain its amplitude down to redshifts $z \sim
2$. Determining the amplitude at lower redshifts is, however,
challenging. Weak gravitational lensing of galaxies is a particularly promising
technique.  Several weak lensing
analyses \citep{Heymans:2013,cfhtlens,discrep4, kids, des} have
reported constraints on the amplitude of the fluctuation spectrum as
measured by the parameter $\sigma_8$.\footnote{Here, $\sigma_8$ is the
  rms amplitude of the mass fluctuations in spheres of radius $8
  h^{-1}\,{\rm Mpc}$, where $h$ is the Hubble constant in units of $100\,
  {\rm km}\, {\rm s}^{-1}\, {\rm Mpc}^{-1}$.}  The results from two surveys, 
CFHTLenS \citep{discrep4} and KiDS \citep{kids}, are discrepant with
the Planck constraints on the parameter combination
$\sigma_8\Omega_m^{0.5}$, where $\Omega_m$ is the present-day matter
density parameter, at about the $2.5\, \sigma$ level, assuming the
standard six-parameter $\Lambda$CDM cosmology (which we will refer to
as the base-$\Lambda$CDM model). The possibility of new physics beyond
base-$\Lambda$CDM \citep{evde} merits close scrutiny of both the weak
lensing and CMB data.

Recently, Ref.~\citep{limits} investigated the Limber
approximation \citep{limber} and a number of other small-angle
approximations used to relate weak lensing observables to the
three-dimensional matter power spectrum. The first version of that
paper concluded that such approximations could contribute
significantly to the tension between the CMB measurements and weak
lensing data. This conclusion, if correct, would have important
implications for cosmology and motivated the analysis presented in
this paper.

The paper is structured as follows. In Sec.~\ref{sec:tps} we derive the full
two-point statistics for weak galaxy lensing in different tomographic
redshift bins in full generality, i.e., without using flat-sky or
Limber-like approximations (extending the work of
Refs~\citep{ext_lim,ext_lim3}). We then compare the exact formulae to
small-angle approximations.  Section~\ref{sec:data} applies these formulae to the latest
tomographic weak lensing data from CFHTLenS\footnote{The data used in this paper is publicly available at \url{https://github.com/sjoudaki/cfhtlens_revisited}.} \citep{discrep4} and
assesses the impact of these approximations on cosmological parameters
and on the tension with the base-$\Lambda$CDM constraints from Planck. Our
conclusions are presented in Sec.~\ref{sec:conc}. Appendix~\ref{sec:relations} provides more detail
on some of the results used in the main text.

\section{Two-point statistics}
\label{sec:tps}

Our aim in this section is to present exact expressions for 
two-point weak lensing statistics and then to apply small-angle
approximations. Further details of the calculations that underlie
these results are given in the appendix. Throughout, we assume a
spatially-flat universe.

We start from the definition of the lensing potential $\phi(\hat{\bm{n}})$ as a
function of the gravitational potential $\Phi(\bm{x},\chi)$ at
(comoving) position $\bm{x}$ and lookback time $\chi$, integrated along the
line-of-sight $\hat{\bm{n}}$ and weighted by a redshift distribution $n(\chi)$ of lensing sources:
\begin{equation}
\label{lenspot}
\phi (\hat{\bm{n}}) = \frac{2}{c^2} \int \frac{\rmd \chi}{\chi} q(\chi) \Phi (\chi\hat{\bm{n}}, \chi),
\end{equation}
where $\chi$ is the comoving radial distance. In this equation $n(\chi)$ is normalized so that $\int n(\chi) \,d\chi = 1$
and  $q(\chi)$ is the lensing efficiency:
\begin{equation}
\label{q}
q(\chi) = \int_\chi^{\chi_H}  \rmd \chi^\prime\, \frac{\chi^\prime - \chi}{\chi^\prime} n(\chi^\prime),
\end{equation}
where $\chi_H$ is the distance to the particle horizon.

We are interested in the two-point statistic
\begin{equation}
\left\langle \phi(\hat{\bm{n}}) \phi(\hat{\bm{n}}') \right\rangle = \left( \frac{2}{c^2} \right)^2 \int \frac{\rmd \chi}{\chi} q(\chi) \int \frac{\rmd \chi'}{\chi'} q(\chi') \left\langle \Phi (\chi\hat{\bm{n}}, \chi) \Phi (\chi'\hat{\bm{n}}', \chi') \right\rangle.
\end{equation}
Expanding the two-dimensional lensing potential $\phi$ in spherical
harmonics, and the gravitational potential in Fourier modes, and using
Poisson's equation, we can
relate the lensing angular power spectrum $C^{\phi\phi}_\ell$, 
\begin{equation}
\left\langle \phi_{\ell m} \phi_{\ell ' m'}  \right\rangle = C_{\ell}^{\phi \phi} \delta_{\ell \ell '} \delta_{m m'}, 
\end{equation}
to the unequal-time dimensional matter power spectrum
$P_\delta(k;\chi,\chi')$. If we further approximate the unequal-time
power spectrum as separable (which is exact in linear theory; see~\cite{2016arXiv161200770K} for
the impact of non-linear evolution), i.e.,
\begin{equation}
P_\delta(k;\chi,\chi') \approx \left[P_\delta(k;\chi)
  P_\delta(k;\chi')\right]^{1/2} ,
\end{equation}
for two tomographic redshift bins ($r, s$) with redshift distributions $n^r(\chi)$ and $n^s(\chi)$ we
find
\begin{subequations}
\begin{equation}
\label{final}
C_{\ell}^{\phi \phi} (r, s) = \frac{8}{\pi} \left(
  \frac{3\Omega_m H_0^2}{2c^2}\right)^2 \int \frac{\rmd k}{k^2} \, I_{\ell}^{r} (k) I_{\ell}^{s} (k),
\end{equation}
where
\begin{equation}
\label{finali}
I_{\ell}^r (k) = \int \frac{\rmd \chi}{\chi} \, \left[ 1 + z(\chi) \right] q^r(\chi) j_{\ell} (k \chi) \left[ P_{\delta}(k;\chi) \right]^{1/2}.
\end{equation}
\end{subequations}
The main observable in weak galaxy lensing surveys is the shear, which
can be described by the spin-2 field $\gamma = \eth^2 \phi /2$, where
$\eth$ is the spin-raising operator~\cite{spin-weight}. The expansions
of the shear, and its complex conjugate, in spin $\pm 2$ spherical harmonics are
\begin{align}
\gamma(\hat{\bm{n}}) &= \sum_{\ell m} (\epsilon_{\ell m} + i
                       \beta_{\ell m}) {}_2
                       Y_{\ell m}(\hat{\bm{n}}) , \\
\gamma^\ast(\hat{\bm{n}}) &= \sum_{\ell m} (\epsilon_{\ell m} - i
                            \beta_{\ell m}) {}_{-2}
                       Y_{\ell m}(\hat{\bm{n}}) , 
\end{align}
where $\epsilon_{\ell m}$ are the multipoles of the E-mode of the shear
and $\beta_{\ell m}$ for the B-modes. For lensing in the Born
approximation, as considered here, the gravitational shear has only
E-modes with $\epsilon_{\ell m} = \sqrt{(\ell+2)!/(\ell-2)!}\phi_{\ell m}/2$.
The angular power spectrum of the lensing potential,
Eq.~\eqref{final}, can easily be related to the E-mode  power spectrum
of the shear field by
\begin{equation}
\label{shear}
C_{\ell}^{\epsilon\epsilon} = \frac{1}{4} \frac{(\ell+2)!}{(\ell-2)!} C_{\ell}^{\phi \phi}.
\end{equation}

The shear $\gamma$ is defined relative to the $\theta$ and $\phi$
directions. The two-point functions of the shear in real space are
simplest when the shear at the two points, $\hat{\bm{n}}_1$ and
$\hat{\bm{n}}_2$, are rotated onto bases generated by the great circle
through the two points (and the orthogonal directions). If we denote
the rotated shear by an overbar, e.g., $\bar{\gamma}(\hat{\bm{n}}_1)$,
the shear correlation functions for tomographic bins $r$ and $s$ are
defined as
\begin{equation}
\xi_+ (\theta; r, s) = \langle \bar{\gamma}_r^\ast(\hat{\bm{n}}_1)
\bar{\gamma}_s(\hat{\bm{n}}_2) \rangle \, , \qquad
\xi_- (\theta; r, s) = \langle \bar{\gamma}_r(\hat{\bm{n}}_1)
\bar{\gamma}_s(\hat{\bm{n}}_2) \rangle ,
\end{equation}
where $\theta$ is the angle between $\hat{\bm{n}}_1$ and
$\hat{\bm{n}}_2$.
The $\xi_{\pm}$ can be expressed in terms of the shear power spectrum
as\footnote{Equation~(48) of
    Ref.~\cite{Kilbinger:2017} incorrectly has the $d^\ell_{2\, 2}$ replaced by Legendre polynomials in the expression for the
  spherical $\xi_+(\theta;r,s)$. This introduces errors that increase steadily with $\theta$. However, for our application to CFHTLenS (Sec.~\ref{sec:data}) the error is below $2\,\%$ on the relevant scales ($\theta < 100\,\mathrm{arcmin}$) and so does not affect the conclusions of Ref.~\cite{Kilbinger:2017}.}
\begin{subequations}
\begin{align}
\label{xip}
\xi_+ (\theta; r, s) & = \sum_{\ell} \frac{2 \ell + 1}{4
                       \pi}  C_{\ell}^{\epsilon\epsilon} (r, s) d^l_{2\, 2} (\theta), \\
\label{xim}
\xi_- (\theta; r, s) &= \sum_{\ell} \frac{2 \ell + 1}{4
                       \pi} C_{\ell}^{\epsilon\epsilon} (r, s) d^{\ell}_{2\, -2} (\theta), 
\end{align}
\end{subequations}
where $d^{\ell}_{mn}$ are the reduced Wigner $D$-matrices.
The equations above are all exact and make no use of the Limber or flat-sky approximations.

\subsection{Limber approximation}
\label{sec:limber}

The exact expressions for the power spectra, Eqs.~(\ref{final})
and~(\ref{finali}) are time consuming to evaluate accurately at high
multipoles due to the rapid oscillations of the spherical Bessel
functions. Many analyses adopt the Limber approximation instead, which
is accurate at large $\ell$ and much easier to compute. In the Limber
approximation, we effectively replace the spherical Bessel function in
Eq.~(\ref{finali}) with a delta-function,
\begin{equation}
j_{\ell} (k \chi) \rightarrow \sqrt{\frac{\pi}{2 \nu}} \delta_D (\nu -
k \chi) ,
\end{equation}
where $\nu = \ell + 1/2$. The wavenumber $k$ is then related to radial distance $\chi$ via 
the relation $k \chi = \nu$. This approximation is accurate if the
rest of the integrand in Eq.~(\ref{finali}) is slowly varying compared to the spherical
Bessel function, which is generally the case at large $\ell$
(see~\cite{Loverde2008} for a careful discussion). Making this
approximation in Eq.~(\ref{finali}), and using Eqs.~\eqref{final}
and~\eqref{shear}, we recover the familiar Limber formula for the
shear power spectrum
\begin{equation}
\label{lim_ps}
C_{\ell}^{\epsilon\epsilon}(r, s) =  \frac{(\ell+2)!}{\nu^4 (\ell-2)!}
\left(\frac{3\Omega_m H_0^2}{2c^2} \Omega_m \right)^2 \int 
\rmd \chi \left[1 + z(\chi) \right]^2 q^r(\chi) q^s(\chi)
P_{\delta}(\nu/\chi; \chi). 
\end{equation}

\subsection{Flat-sky approximations}
\label{sec:flat-sky}

Mathematically, the flat-sky approximations consist of replacing the expansion in spherical harmonics by an expansion in Fourier modes. The
relation between shear and lensing power spectra, Eq.~\eqref{shear}, is then 
\begin{equation}
\label{shear_flat}
C_{\ell}^{\epsilon\epsilon} \approx \frac{\ell^4}{4} C_{\ell}^{\phi \phi}.
\end{equation}
The reduced $D$-matrices  for high multipoles can be approximated by Bessel functions \citep{corr_flat}:
\begin{equation}
d^{\ell}_{2 ~ +2} (\theta) \approx J_0(\ell \theta) \qquad d^{\ell}_{2 ~ -2} (\theta) \approx J_4(\ell \theta),
\end{equation}
and together  with the Limber-approximated expression \eqref{lim_ps}, we obtain the usual expression for the shear power spectrum 
\citep{standard_ps}
\begin{subequations}
\begin{equation}
\label{lim_ps_flat}
C_{\ell}^{\epsilon\epsilon}(r, s) = \left(\frac{3\Omega_m H_0^2}{2c^2} \right)^2 
\int \rmd \chi \, \left[ 1 + z(\chi) \right]^2 q^r(\chi) q^s(\chi)
P_{\delta}(\nu/\chi; \chi), 
\end{equation}
and the correlation functions
\begin{eqnarray}
\label{xip_flat}
\xi_+ (\theta, r, s) = \frac{1}{2 \pi} \int \rmd \ell \ell J_0  (\ell \theta) C_{\ell}^{\epsilon\epsilon} (r, s), \\
\label{xim_flat}
\xi_- (\theta, r, s) = \frac{1}{2 \pi} \int \rmd \ell \ell J_4  (\ell \theta) C_{\ell}^{\epsilon\epsilon} (r, s).
\end{eqnarray}
\end{subequations}
Note that we have replaced $\nu$ by $\ell$ in the prefactor of the
Limber-approximated power spectrum, Eq.~\eqref{lim_ps}, and also in
the expressions~\eqref{xip} and~\eqref{xim} for the correlation
functions [$(2\ell+1)/(4\pi) \rightarrow \ell/(2\pi)$]. We have,
however, retained $\nu$ in the argument of the matter power spectrum.

In this paper, we compare exact results with two flat-sky
approximations: (i) Equations~(\ref{lim_ps_flat}--\ref{xim_flat}) with $\nu
= (\ell + 1/2)$ in the argument of the matter power spectrum,
which is the approximation used in the tomographic  analysis of
CFHTlenS and KiDS weak lensing data\footnote{Note that Eq.~(4) in
  \citep{kids} uses $\nu = \ell$ in the argument of the matter power
  spectrum; however, the public KiDS likelihood code uses $\nu = \ell
  + 1/2$.} (we call this `flat-sky no prefactor'); and 
(ii) Equations~\eqref{lim_ps}, and (\ref{xip_flat}--\ref{xim_flat}) with
$\nu = (\ell + 1/2)$ in both the prefactor and the matter power
spectrum (which we call `flat-sky with prefactor'). These two
approximations differ from each other only in the prefactor of the
shear power spectrum. The spectrum with the prefactor is smaller by a
factor of approximately $1-5/(2\ell^2)$.
We shall compare each 
of these approximations to the exact expressions~(\ref{final}--\ref{finali}) and~(\ref{xip}--\ref{xim}).

\section{Application to  CFHTLenS data}
\label{sec:data}

\begin{figure}
\centering
\centering
\includegraphics[width=.49\textwidth]{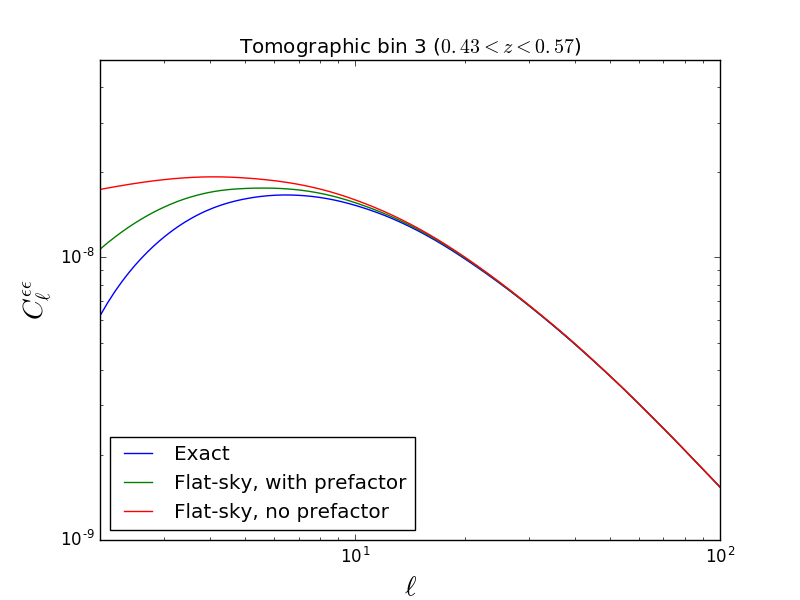}
\includegraphics[width=.49\textwidth,origin=c]{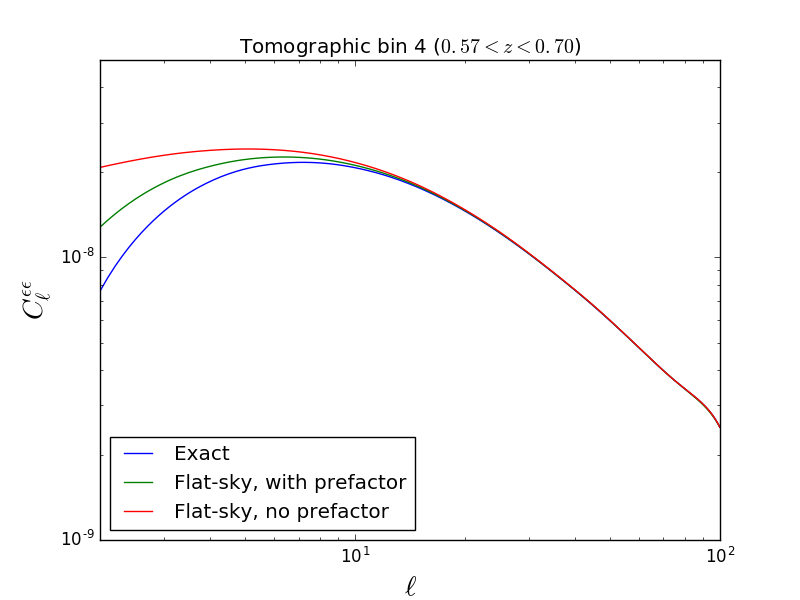}
\caption{\label{fig:cl3} Shear power spectra for the third (left) and
  fourth (right) tomographic redshift bins of the revised CFHTLenS data set
assuming the best-fit parameters of \citep{discrep4}, 
illustrating  the effects of the small-angle approximations discussed
in the text. The red lines use the same approximations as in the
tomographic analyses of the CFHTlenS and KiDS weak lensing data, while
the blue lines are from the exact calculation.
}
\end{figure}

In this section we investigate the impact of the small-angle
approximations on current weak lensing parameter constraints using the
CFHTLenS survey as an example.  CFHTLenS is based on imaging data
from the Canada-France-Hawaii telescope in five photometric bands. In the 
reanalysis of \citep{discrep4}, the data is divided
into seven tomographic redshift bins with photometric redshifts
spanning the range  $0.15< z < 1.30$. We use the calibrated redshift
distributions to compute the exact shear power spectrum, as well as
the `flat-sky no prefactor' and the `flat-sky with prefactor' spectra,
using the formulae presented in the previous section assuming the best-fit parameters 
of \citep{discrep4}.  Figure~\ref{fig:cl3} shows the resulting shear power spectra for the third and fourth 
tomographic redshift  bins. This shows that the small-angle approximations have 
negligible impact except at  multipoles $\ell \simlt 10$. However, the CFHTLenS 
analyses are insensitive to these multipoles.

\begin{figure}
\centering
\includegraphics[width=.49\textwidth]{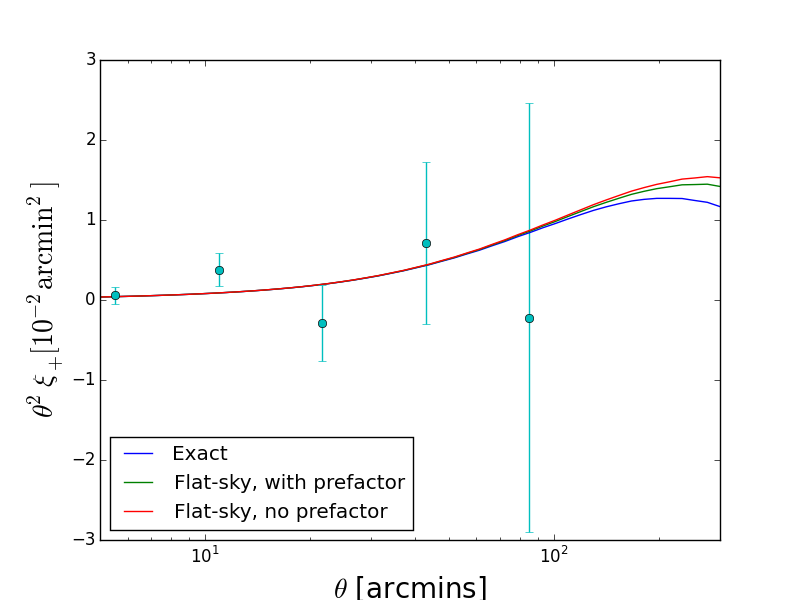}
\includegraphics[width=.49\textwidth,origin=c]{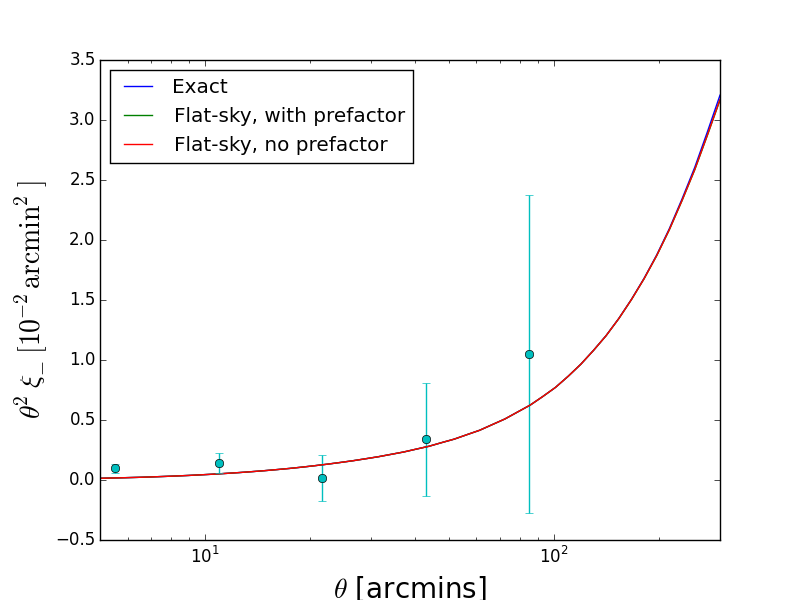}
\caption{\label{fig:xi3} Shear correlation functions $\xi_+$ (left) and $\xi_-$ (right) for 
the third tomographic redshift bin of CFHTLenS. The points show the CFHTLenS measurements together with $1\,\sigma$ errors.
The model curves show the effects of the small-angle approximations described in the text, assuming the best-fit parameters
of~\citep{discrep4}.}
\end{figure}

We can see explicitly that the small-angle approximations have
negligible impact on the CFHTLens analysis by looking at the shear correlation functions
$\xi_+$ and $\xi_-$. Figure~\ref{fig:xi3} shows the predictions for $\xi_+$ and
$\xi_-$ for the various approximations assuming the best-fit parameters of
\citep{discrep4} for tomographic redshift bin 3, together with the data
points and $1\,\sigma$ errors from \citep{discrep4}. Compared to the large
errors on the CFHTLenS data points, the effects of adopting small-angle approximations
are negligible over the angular scales probed by the data.

As a final test, we perform parameter estimation (with CosmoMC
\citep{cosmomc,cosmomc2}) sampling  the CFHTlenS
likelihood as in \citep{discrep4} comparing the `flat-sky prefactor' with the
`flat-sky no prefactor' approximations.\footnote{We have not tested the exact calculation
since it is too slow to be used in CosmoMC. However, the differences
between the exact correlation functions and those with the `flat-sky
no prefactor' approximation are about twice as large as the differences
between the two flat-sky approximations, so our analysis should still be
representative of the actual errors introduced by the
small-angle approximations.} As expected from Fig. \ref{fig:cl3}, the impact 
of these approximations on cosmological parameters such as $\sigma_8$ and $\Omega_m$ is
undetectable within the convergence errors of the parameter chains (and therefore well below
the $1\,\sigma$ errors on cosmological parameters).

\section{Conclusions}
\label{sec:conc}

The analysis presented here shows that the small-angle Limber and
flat-sky approximations that are typically used in the analysis of
galaxy weak lensing have no significant impact on cosmological
parameters derived from  current data.
Kilbinger et al.~\citep{Kilbinger:2017} have reached similar conclusions recently, as does the revised version of \citep{limits}. Three
independent analyses are therefore in agreement that small-angle
approximations have no bearing on the discrepancy between the
amplitude of the mass fluctuation spectrum inferred from the CFHTLenS
or KiDS galaxy weak lensing data and that measured
from the CMB assuming the base-$\Lambda$CDM cosmology.

\acknowledgments

Pablo Lemos acknowledges support from an Isaac Newton Studentship at the University of Cambridge and from the Science and Technologies Facilities Council (STFC). We thank Patricia Larsen for helpful discussions.

\appendix

\section{Galaxy weak lensing two-point statistics}
\label{sec:relations}

In this appendix we summarise results for the two-point statistics of the
weak lensing observables, namely convergence and shear, on the
spherical sky and also in the flat-sky limit. These results are valid
for both weak lensing of galaxies and of the CMB. As in the main text,
we assume a flat universe throughout.

All weak lensing quantities can be defined as a function of the
lensing potential $\phi(\hat{\bm{n}},\chi)$ for sources at comoving
distance $\chi$:
\begin{equation}
\phi (\hat{\bm{n}}, \chi) = \frac{2}{c^2} \int_0^{\chi}\rmd \chi'\, \frac{\chi - \chi'}{\chi\chi'} \Phi (\chi'\hat{\bm{n}}, \chi'),
\end{equation}
where $\Phi (\bm{x}, \chi)$ is the gravitational potential at comoving
position $\bm{x}$ and conformal lookback time $\chi$.

If we average observables over sources with a redshift distribution
$n(\chi)$, normalised such that $\int n(\chi) \, \rmd \chi = 1$, the
relevant 2D lensing potential is given by
\begin{equation}
\label{lenspot_ap}
\phi (\hat{\bm{n}}) = \int \rmd \chi \, n(\chi) \phi (\hat{\bm{n}}, \chi) = \frac{2}{c^2} \int \frac{\rmd \chi'}{\chi'} q(\chi') \Phi (\chi'\hat{\bm{n}}, \chi'),
\end{equation}
where $q(\chi)$ is the lensing efficiency defined in Eq.~\eqref{q}. In the case of CMB lensing, the source distribution can be
approximated by a delta-function at the surface of last scattering:
$n(\chi) = \delta_D(\chi - \chi_\ast)$. The lensing potential is a
scalar field and can be expanded in spherical harmonics as
\begin{equation}
\label{sh_phi}
\phi(\hat{\bm{n}}) = \sum_{\ell, m} \phi_{\ell m} Y^{m}_{\ell} (\hat{\bm{n}}).
\end{equation}

The basic observables in weak lensing are the convergence $\kappa$ and
the components $\gamma_1$ and $\gamma_2$ of the shear, defined in
terms of the second derivatives of the lensing potential as
\begin{equation}
\nabla_i \nabla_j \phi = \kappa g_{ij} + \frac{1}{2}(\gamma_1 + i \gamma_2)(\bm{m}_- \otimes \bm{m}_-)_{ij} + \frac{1}{2}(\gamma_1 - i \gamma_2)(\bm{m}_+ \otimes \bm{m}_+)_{ij},
\end{equation}
where $g_{ij}$ is the metric on the sphere and the null vectors $\bm{m}_\pm =
\hat{\bm{\theta}} \pm i \hat{\bm{\phi}}$. Here, $\hat{\bm{\theta}}$ and
  $\hat{\bm{\phi}}$ are unit vectors along the $\theta$ and $\phi$
  coordinate directions of a spherical-polar coordinate system. The
  convergence $\kappa = \nabla^2 \phi /2$ describes isotropic
  magnification/dilation and is a scalar field. The shear describes
  area-preserving distortions; the complex shear $\gamma = \gamma_1  +
  i \gamma_2$ is a spin-2 field related to the lensing potential via
\begin{equation}
\gamma = \frac{1}{2}(\bm{m}_+ \otimes \bm{m}_+)_{ij} \nabla^i \nabla^j \phi =
\frac{1}{2}\eth^2 \phi ,
\end{equation}
where $\eth$ is the spin-raising operator~\cite{spin-weight}. The
spherical-harmonic expansions of the convergence and shear follow from Eq.~\eqref{sh_phi}:
\begin{subequations}
\begin{eqnarray}
\label{kappa}
\kappa (\hat{\bm{n}}) = - \frac{1}{2} \sum_{\ell, m} \ell (\ell + 1) \phi_{\ell m} Y^m_{\ell} (\hat{\bm{n}}), \\
\label{gamma}
\gamma_1 (\hat{\bm{n}}) \pm i \gamma_2 (\hat{\bm{n}}) = \frac{1}{2}
  \sum_{\ell, m}\sqrt{\frac{(\ell + 2)!}{(\ell - 2)!}} \phi_{\ell m} ~ {}_{\pm 2} Y^m_{\ell} (\hat{\bm{n}}).
\end{eqnarray}
\end{subequations}

Generally, a spin 2 field can be expanded in E- and B-modes, for
example,
\begin{equation}
(\gamma_i \pm i \gamma_2)(\hat{\bm{n}}) = \sum_{\ell,m} \left(\epsilon_{\ell m} \pm i
  \beta_{\ell m}\right) {}_{\pm 2}Y_\ell^m(\hat{\bm{n}}) ,
\end{equation}
where $\epsilon_{\ell m}$ are the E-mode multipoles and $\beta_{\ell
  m}$ are the B-mode multipoles. These transform oppositely under
parity: $\epsilon_{\ell m} \rightarrow (-1)^\ell \epsilon_{\ell m}$
and $\beta_{\ell m} \rightarrow (-1)^{\ell+1} \beta_{\ell m}$. However, we see from
Eq.~\eqref{gamma} that the gravitational shear has no B-modes (in the
Born approximation that we are assuming here), while $\epsilon_{\ell
  m} = \sqrt{(\ell+2)!/(\ell-2)!}\phi_{\ell m}/2$.

The angular power spectrum of the lensing potential is defined by
\begin{equation}
\label{cphiphi}
\langle \phi_{\ell m} \phi^\ast_{\ell' m'} \rangle = \delta_{\ell
  \ell'} \delta_{m m'} C_{\ell}^{\phi \phi} .
\end{equation}
We can similarly define the convergence and shear power, which are
related to $C_\ell^{\phi\phi}$ by
\begin{subequations}
\begin{align}
C_{\ell}^{\kappa \kappa} &= \frac{\ell^2(\ell+1)^2}{4} C_{\ell}^{\phi \phi}, \\
\label{shear_ap}
C_{\ell}^{\epsilon\epsilon} &= \frac{1}{4} \frac{(\ell+2)!}{(\ell-2)!} C_{\ell}^{\phi \phi}.
\end{align}
\end{subequations}

In the flat-sky approximation, we project onto the tangent plane at
the centre of the observed field and denote positions in this plane
with a 2D vector $\bm{\theta}$. Expansions in spherical harmonics are
replaced by Fourier expansions, so that for the gravitational
potential
\begin{equation}
\label{fou_phi}
\phi(\bm{\theta}) = \int \frac{\rmd^2 \bm{l}}{(2 \pi)^2}\, \phi
(\bm{l}) e^{i \bm{l} \cdot \bm{\theta}} .
\end{equation}
The expansions of the convergence and shear become
\begin{subequations}
\begin{align}
\label{kappa_fl}
\kappa (\bm{\theta}) &= - \frac{1}{2} \int \frac{\rmd^2 \bm{l}}{(2 \pi)^2}\, l^2 \phi (\bm{l}) e^{i \bm{l} \cdot \bm{\theta}} \\
\label{gamma_fl}
(\gamma_1 \pm i \gamma_2)(\bm{\theta}) &= - \frac{1}{2} \int
  \frac{\rmd^2 \bm{l}}{(2 \pi)^2}\, l^2 \phi (\bm{l}) e^{\pm 2
  i (\psi_{\bm{l}} - \psi_{\bm{\theta}})} e^{i \bm{l} \cdot
                                         \bm{\theta}} ,
\end{align}
\end{subequations}
where $\psi_{\bm{l}}$ and $\psi_{\bm{\theta}}$ are the angles that
$\bm{l}$ and $\bm{\theta}$, respectively, make with the
$x$-axis. Note that the components of the shear are defined relative
to a polar-coordinate basis in the plane; rotating to a global
Cartesian basis removes the factors of $e^{\mp 2 i \psi_{\bm{\theta}}}$.

In the flat-sky approximation, the power spectra are related simply by 

\begin{equation}
C_{\ell}^{\kappa \kappa} = \frac{\ell^4}{4} C_{\ell}^{\phi \phi} = \frac{\ell^4}{4} C_{\ell}^{\epsilon\epsilon}.
\end{equation}

\subsection{Relation to the matter power spectrum}

We can relate the lensing angular power spectra to the 3D matter power
spectra as follows. We begin by expanding the gravitational potential in
Eq.~\eqref{lenspot_ap} in Fourier modes $\Phi(\bm{k},\chi)$ and using the plane-wave
expansion to find
\begin{equation}
\label{eq:Phitophi}
\phi_{\ell m} = i^\ell \frac{8\pi}{c^2} \int \frac{\rmd^3
  \bm{k}}{(2\pi)^3} \left(\int \frac{\rmd \chi}{\chi} \, q(\chi)
j_\ell(k\chi) \Phi(\bm{k},\chi) \right) Y_\ell^m{}^\ast(\hat{\bm{k}})
\, .
\end{equation}
The two-point correlator of the gravitational potential is
\begin{equation}
\langle \Phi(\bm{k},\chi) \Phi^\ast(\bm{k}',\chi') \rangle = (2\pi)^3
P_\Phi(k;\chi,\chi') \delta_D^{(3)}(\bm{k}-\bm{k}') ,
\end{equation}
where $P_\Phi(k;\chi,\chi')$ is the unequal-time power spectrum of the
gravitational potential. It follows from Eq.~\eqref{eq:Phitophi} that,
for redshift distributions $q^r(\chi)$ and $q^s(\chi)$,
\begin{equation}
C_\ell^{\phi\phi}(r,s) = \left(\frac{8\pi}{c^2}\right)^2 \int
\frac{k^2 \rmd k}{(2\pi)^3} \int \frac{\rmd \chi}{\chi} q^r(\chi)
j_\ell(k\chi)
\int \frac{\rmd \chi'}{\chi'} q^s(\chi') j_\ell(k\chi')
P_\Phi(k;\chi,\chi') .
\end{equation}
This simplifies if we adopt the approximation made in the main text,
\begin{equation}
P_\Phi(k;\chi,\chi') \approx \left[P_\Phi(k;\chi)P_\Phi(k;\chi')\right]^{1/2} ,
\end{equation}
to give
\begin{equation}
C_{\ell}^{\phi \phi}(r,s) = \frac{8}{\pi c^4} \int k^2 \rmd k \, \hat{I}^r_{\ell} (k) \hat{I}^s_{\ell} (k)  , 
\end{equation}
where 
\begin{equation}
\hat{I}^r_{\ell} (k) = \int \frac{\rmd \chi}{\chi} q^r(\chi) j_{\ell} (k \chi) \left[ P_{\Phi}(k,\chi) \right]^{1/2}.
\end{equation}
Finally, we can relate the 3D power spectrum of the gravitational
potential to the 3D matter power spectrum using Poisson's equation, i.e.,
\begin{equation}
P_{\Phi}(k;\chi) = \left( \frac{3}{2} \Omega_m H_0^2 \left[1+z(\chi)\right] \right)^2 \frac{P_{\delta}(k;\chi)}{k^4},
\end{equation}
to obtain our final result
\begin{subequations}
\begin{align}
\label{final_ap}
C_{\ell}^{\phi \phi} (r, s) &= \frac{8}{\pi} \left( \frac{3 \Omega_m
  H_0^2}{2 c^2} \right)^2 \int \frac{\rmd k}{k^2}\, I_{\ell}^{r} (k) I_{\ell}^{s} (k), \\
\label{finali_ap}
I_{\ell}^r (k) &= \int \frac{\rmd \chi}{\chi} \left[ 1 + z(\chi) \right] q^r(\chi) j_{\ell} (k \chi) \left[ P_{\delta}(k,\chi) \right]^{1/2}.
\end{align}
\end{subequations}

\subsection{Spherical correlation functions}

Expressions for the spherical correlation functions of the
gravitational shear from the angular power spectrum can be obtained
following the methods used for CMB polarization (also a spin-2 field)
in Refs.~\cite{pol,pol2}. To maintain generality, we give results
including B-modes although, as noted above, these are expected to vanish for
the gravitational shear.

As discussed in the main text, the correlation functions of the shear
for lines of sight $\hat{\bm{n}}_1$ and $\hat{\bm{n}}_2$ are simplest
when the shear is expressed at each point in bases generated by the
great circle through the two points. If $\alpha_1$ is the angle
require to rotate $\hat{\bm{\theta}}_1$ in a right-handed sense about
$\hat{\bm{n}}_1$ onto the tangent to the great
circle there, the rotated shear is $\bar{\gamma}(\hat{\bm{n}}_1) =
e^{-2i\alpha_1} \gamma(\hat{\bm{n}}_1)$. For redshift distributions
$q^r(\chi)$ and $q^s(\chi)$, the two-point correlation functions of
the rotated shear are
\begin{subequations}
\begin{align}
\label{xip_app}
\xi_+ (\theta; r, s) & = \langle \bar{\gamma}_r^\ast(\hat{\bm{n}}_1)
                       \bar{\gamma}_s(\hat{\bm{n}}_2)\rangle = \sum_{\ell} \frac{2 \ell + 1}{4
                       \pi}  \left[C_{\ell}^{\epsilon\epsilon} (r, s)
                       + C_\ell^{\beta\beta}(r,s)\right] d^l_{2\, 2} (\theta), \\
\label{xim_app}
\xi_- (\theta; r, s) &= \langle \bar{\gamma}_r(\hat{\bm{n}}_1)
                       \bar{\gamma}_s(\hat{\bm{n}}_2)\rangle = \sum_{\ell} \frac{2 \ell + 1}{4
                       \pi} \left[ C_{\ell}^{\epsilon\epsilon} (r, s) - C_\ell^{\beta\beta}(r,s)
                       \right] d^{\ell}_{2\, -2} (\theta), 
\end{align}
\end{subequations}
where $d^{\ell}_{mn}$ are the reduced Wigner $D$-matrices, and
$\theta$ is the angle between $\hat{\bm{n}}_1$ and $\hat{\bm{n}}_2$.
The correlation coefficients for the shear components follow from
these expressions (noting that the right-hand sides are real-valued):
\begin{subequations}
\begin{align}
\langle \bar{\gamma}_{1,r}(\hat{\bm{n}}_1)
  \bar{\gamma}_{1,s}(\hat{\bm{n}}_2) \rangle &=
                                               \frac{1}{2}\left[\xi_+(\theta;r,s)
                                               +
                                               \xi_-(\theta;r,s)\right]
                                               , \\
\langle \bar{\gamma}_{2,r}(\hat{\bm{n}}_1)
  \bar{\gamma}_{2,s}(\hat{\bm{n}}_2) \rangle &=
                                               \frac{1}{2}\left[\xi_+(\theta;r,s)
                                               -
                                               \xi_-(\theta;r,s)\right]
                                               .
\end{align}
\end{subequations}
Correlations between mixed components, e.g., $\langle \bar{\gamma}_{1,r}(\hat{\bm{n}}_1)
  \bar{\gamma}_{2,s}(\hat{\bm{n}}_2) \rangle$, vanish since we are
  assuming that parity invariance holds in the mean (so that
  $C_\ell^{\epsilon\beta}(r,s) = 0$).




\bibliographystyle{JHEP}

\bibliography{refs}

\end{document}